%% file: main.tex
\documentclass[12pt]{article}

\usepackage{authblk}

\title{\textbf{Single-Pixel Multimode Fiber Spectrometer via Wavefront Shaping}}

\author[1, 2]{Şahin Kürekci \thanks{Corresponding author: kurekci@metu.edu.tr} {}}
\author[1, 3]{S. Süleyman Kahraman}
\author[1, 2]{Emre Yüce}

\affil[1]{Programmable Photonics Group, Department of Physics, Middle East Technical University, Ankara, Turkey}
\affil[2]{The Center for Solar Energy Research and Applications (ODTÜ-GÜNAM), Middle East Technical University, Ankara, Turkey}
\affil[3]{Caltech Optical Imaging Laboratory, Andrew and Peggy Cherng Department of Medical Engineering, Department of Electrical Engineering, California Institute of Technology, Pasadena, CA, USA}

\date{}                     
\include{packages}

\begin{document}

\maketitle


\begin{abstract}
When light passes through a multimode fiber, two-dimensional random intensity patterns are formed due to the complex interference within the fiber. The extreme sensitivity of speckle patterns to the frequency of light paved the way for high-resolution multimode fiber spectrometers. However, this approach requires expensive IR cameras and impedes the integration of spectrometers on-chip. In this study, we propose a single-pixel multimode fiber spectrometer by exploiting wavefront shaping. The input light is structured with the help of a spatial light modulator, and optimal phase masks, focusing light at the distal end of the fiber, are stored for each wavelength. Variation of the intensity in the focused region is recorded by scanning all wavelengths under fixed optimal masks. Based on the intensity measurements, we show that an arbitrary input spectrum having two wavelengths 20 pm apart from each other can be reconstructed successfully (with a reconstruction error of $\sim$3\%) in the near-infrared regime, corresponding to a resolving power of $ R \approx 10^5 $. We also demonstrate the reconstruction of broadband continuous spectra for various bandwidths. With the installation of a single-pixel detector, our method provides low-budget and compact detection at an increased single-to-noise ratio.
\end{abstract}

\newpage

\section*{Introduction}

Spectrometers achieve spectral-to-spatial mapping which allows spectral decomposition of the input light. Conventional spectrometers use dispersive media such as gratings or prisms, and they can achieve resolving powers around $R<10^4$, and can reach up to $R<10^5$ only with the installation of complex triple-grating systems (the spectral resolution is $\delta\lambda = \lambda_0/R$, where $\lambda_0$ is the operating wavelength). However, such spectrometers generally require moving parts (grating, mirrors) and a line array detector for scanning whole wavelengths of interest. Moreover, inverse proportionality between spectral resolution and optical path length leads to bulky systems when high-resolution is demanded. The fundamental need for high resolution spectral analysis in various lines of research and applications triggers new concepts that are built on the basis of holography \cite{patent-1}, scattering of light by a photonic crystal  \cite{Xu2003, Redding2013-1}, a random scattering medium  \cite{Kohlgraf-Owens2010}, or a multimode fiber (MMF) \cite{patent-2, Redding2013, Redding2014, Wan2015} to form a complex spatial intensity distribution (a speckle pattern) on a multipixel detector such as a charged couple device (CCD) or a focal plane array (FPA). In such systems, wavelengths experience different propagation constants inside the scattering medium, thus forming distinct spatial intensity profiles on the detector, which provides the required one-to-one spectral-to-spatial mapping. Before the use of spectrometer, a calibration matrix is measured by scanning all wavelengths in the operational range and it stores the corresponding speckle patterns. The calibration matrix is then utilized to reconstruct an arbitrary input spectrum based on the measured intensity distribution. However, the increased cost of CCD and FPA sensors especially in infrared regime limits the deployment of high-resolution spectral analysis tools.

Among all scattering-based systems, the multimode fiber spectrometers have been particularly attractive by offering high-resolutions with reduced scattering losses (keeping the light collimated inside the fiber and preventing scattering to higher angles). Since fibers can be wrapped, higher spectral resolution can be achieved without enlarging the system. It was shown in \cite{Redding2014} that high resolving powers $R>10^6$ in near-infrared regime is possible with fibers of 100 m long. Yet the signal-to-noise ratio (SNR) is the main limiting factor for the resolution at low signal levels and increased fiber length \cite{Cao2017}.

Single pixel detection together with compressed sensing \cite{Takhar2006, Duarte2008, Gattinger2021} have been revolutionizing imaging methods. Surprisingly, the penetration of this methods in spectroscopy have been very limited due to mechanical resolution limits. The single-pixel imaging (SPI) systems are based on the use of a spatial light modulator (SLM) and a single-pixel detector. Employing the SPI method is particularly useful when working in the infrared regime since FPAs get extremely expensive at longer wavelengths \cite{Edgar2019}.

In this paper, we develop a high-resolution single-pixel multimode fiber spectrometer and demonstrate its ability to reconstruct arbitrary spectra. The single-pixel detection is achieved by focusing light on a selected target region of a focal plane array which is employed as a bucket detector. The input wavefronts are structured using a spatial light modulator which provides distinct output intensities at the detector \cite{Vellekoop2007, Vellekoop2008-1, Mosk2012, Vellekoop2015, Horstmeyer2015}. The intensity variations at the target position as a function of input wavelength are used to reconstruct the spectra at a resolution of 20 pm. The increased intensity at the focused point also increases SNR which removes low signal barrier in reaching high resolutions at low signal levels. This, to best of our knowledge, is the first demonstration of a high-resolution scattering medium based spectrometer exploiting single-pixel detection. Replacing an FPA with a single-pixel detector reduces the cost in infrared applications enormously and it also provides a new method for on-chip hyperspectral compressed imaging which is brought by compact size of a single pixel spectrometer.

\section*{Methods}

\begin{figure}[!ht]
\centering \includegraphics[width=0.8\linewidth]{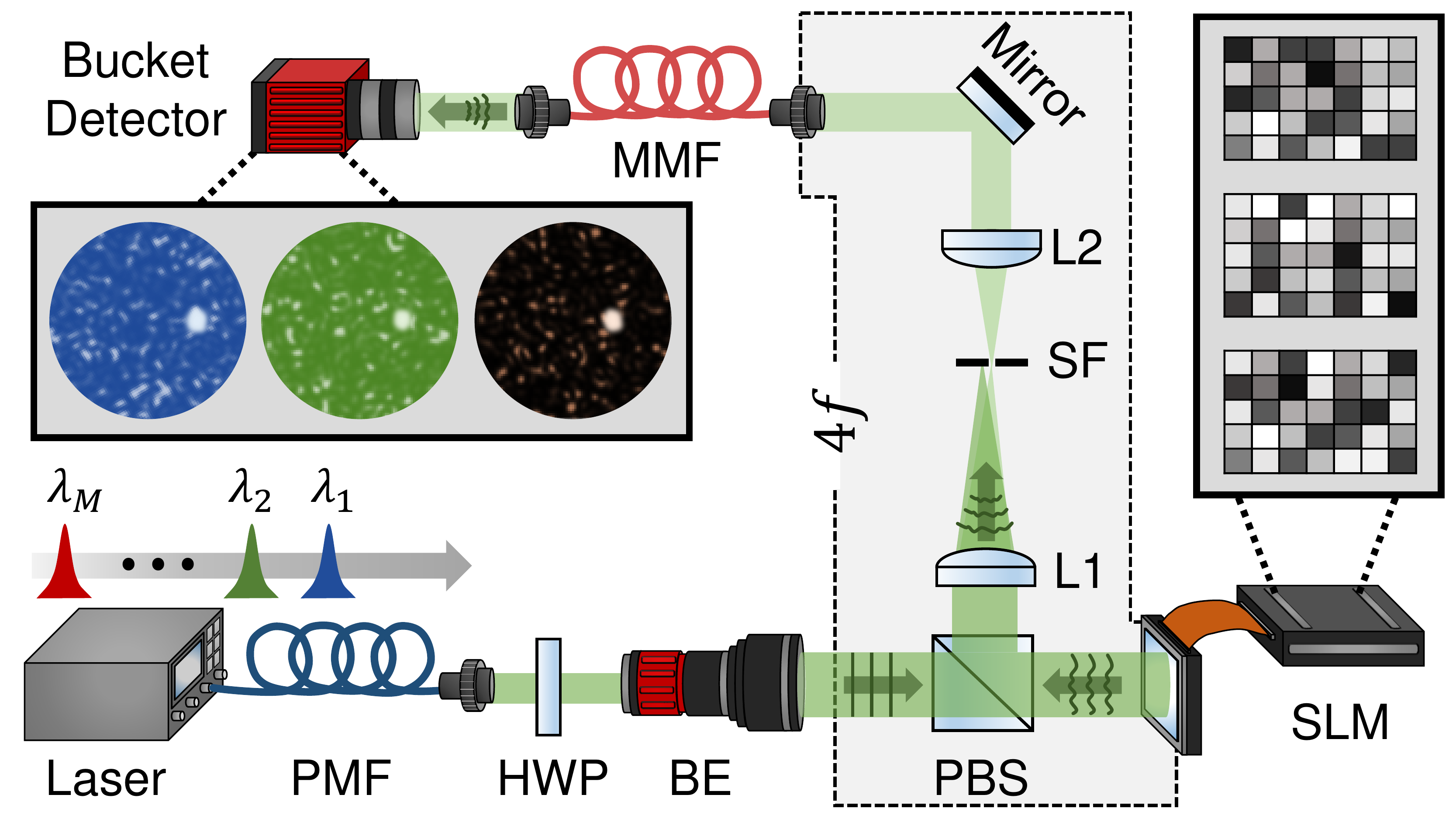}
\caption{Schematic of the setup. The input wavelengths provided by a tunable laser are optimized at the end of a 20-meter-long multimode fiber by using a spatial light modulator. Unique SLM patterns are used to focus distinct wavelengths. L: Lens, PMF: Polarization Maintaining Fiber, HWP: Half-Wave Plate, BE: Beam Expander, PBS: Polarizing Beam Splitter, SLM: Spatial Light Modulator, SF: Spatial Filter, MMF: Multimode Fiber.}
\label{fig:setup}
\end{figure}

The spectrometer is built based on the system given in Fig. \ref{fig:setup}. A tunable laser operating around 1550 nm with 38 nm tuning range is used for illuminating an SLM (HOLOEYE PLUTO-TELCO) with 1920×1080 pixels screen resolution. The beam is expanded to cover the SLM screen, and a half-wave plate is used to control the incident polarization so that it matches the alignment direction of the liquid crystals inside the SLM. To avoid the contribution of the pixelated nature of the SLM on the measurements, a fixed blazed grating is kept on the SLM for all measurements and zeroth order diffracted light is eliminated with a spatial filter \cite{Qi2016}.

\vspace{3mm}

The light is phase-modulated via SLM and focused on the distal end of a 20-meter-long multimode fiber (core diameter 105 µm, NA = 0.22). A continuous sequential algorithm \cite{Vellekoop2008-2} is used in the modulation steps where the SLM is divided into 144 superpixels of size 16×9 and four phase steps between 0-2$\pi$ are scanned through each superpixel. The modulated beam is focused into a circular target region consisting of 76 pixels on a bucket detector of 30 µm×30 µm cell size. The intensity at the focus spot is then integrated to reconstruct the input spectrum from a single-pixel data. Since the whole spectrometer is built upon the intensity measurements of the target region, which effectively covers an area around 0.07 mm$^2$, the bucket detector can safely be replaced with commercial photodiodes which generally have active areas larger than the size of our target region \cite{Donati2021}.

\begin{figure*}[!ht]
\centering \includegraphics[width=\textwidth]{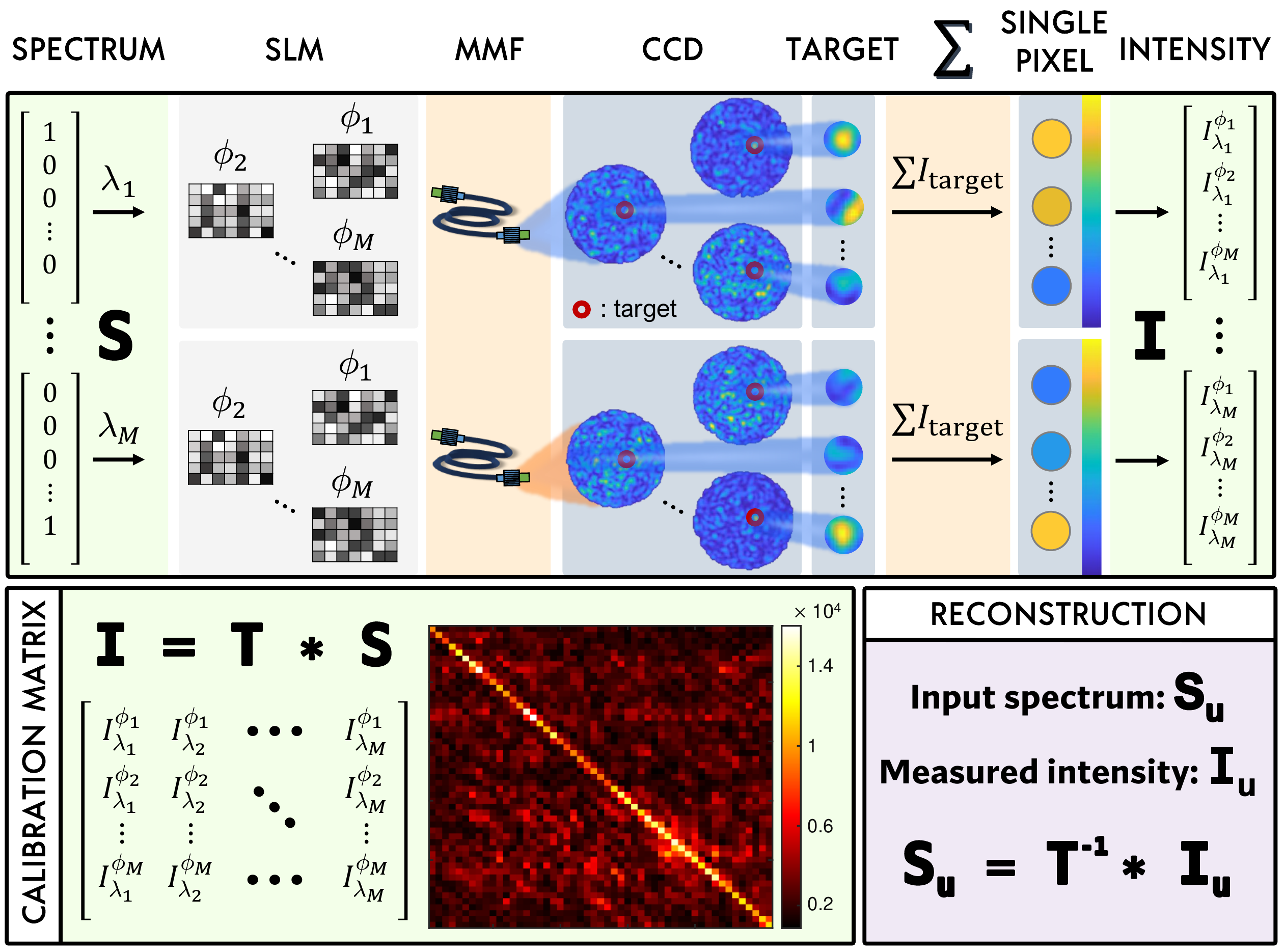}
\caption{Schematic explanation of calibration and reconstruction. The upper rectangle illustrates the process used to obtain the calibration matrix  $\mathbf{T}$. The heatmap of the $50\times50$ calibration matrix obtained in the experiment is shown inside the bottom left rectangle. Using the inverse of the calibration matrix, an unknown input spectrum can be reconstructed as formulated in the bottom right corner. The asterisk operator (*) denotes matrix multiplication.}
\label{fig:calibration}
\end{figure*}

In order to calibrate the spectrometer, we have modified the calibration process in \cite{Redding2013} for an SLM-based single-pixel system as depicted in Fig. \ref{fig:calibration}. We represent the input spectrum of a single wavelength $\lambda_i$ as a unit vector,
\begin{equation}
\mathbf{S_i} = [ \lambda_1 \quad \lambda_2 \quad \ldots \quad \lambda_M]^T,
\label{eq:1}
\end{equation}
where all the elements except $\lambda_i = 1$ are set to zero ($T$ is the transpose operator). All $M$ input wavelengths within the scope of the operating range of the spectrometer are modulated by the SLM and optimized on the target region of the detector. The optimized SLM phase masks, $\phi_i \; (i=1, 2, \ldots, M)$ are stored. Once the optimization process is completed, the target region intensities of individual wavelengths are measured under all recorded SLM patterns, and the intensity values are then integrated to a single numerical value. Throughout the text, $I_{\lambda_i}^{\phi_j}$ denotes  the integrated intensity of the target region under input wavelength $\lambda_i$ and optimized SLM pattern $\phi_j$. For each wavelength $\lambda_i$ (or spectrum vector $\mathbf{S_i}$), a corresponding intensity vector as a function of SLM phase mask $\phi$ is created,
\begin{equation}
\mathbf{I_i} = [ I_{\lambda_i}^{\phi_1} \quad I_{\lambda_i}^{\phi_2} \quad \ldots \quad I_{\lambda_i}^{\phi_M}]^T.
\label{eq:2}
\end{equation}

The intensity vectors are then combined into a single $M \times M$ matrix $\mathbf{T}$, which is called the calibration matrix, whose rows can be used as a measure of system response to wavelength variations. With the implementation of advanced wavefront optimization methods, it is possible to focus light in milliseconds \cite{Xia2019} and complete the calibration of the spectrometer in less than a few minutes for many practical spectroscopy applications.

The entries sitting in the main diagonal of the calibration matrix are expected to have the highest numerical values in their row and column since they correspond to the perfect match between the input wavelength and the optimized SLM pattern. In the off-diagonal entries, wavelength and the optimized phase mask mismatch, and as a result, the integrated intensity drops for steps out of the diagonal. More contrast between the diagonal and off-diagonal entries can be produced when the light is sharply focused on the detector. In this case, the spectrometer is expected to be more sensitive to wavelength variations, thus have a higher resolution. The sharpness of the focus can be controlled by involving a measure of merit (such as enhancement factor) during the optimization. In our experiments, we measured an average enhancement around $\eta \approx 70$, which is approximately 40\% below the theoretical maximum value \cite{Vellekoop2015}.

The calibration process of the spectrometer can be modeled mathematically as a set of linear propagations where the inputs are the spectrum vectors $\mathbf{S_i}$, outputs are the intensity vectors $\mathbf{I_i}$, and the propagation operator is the calibration matrix $\mathbf{T}$,
\begin{equation}
\mathbf{T} * \mathbf{S_i} = \mathbf{I_i}.
\label{eq:3}
\end{equation}
To test the performance of the spectrometer, an unknown spectrum $\mathbf{S_u^{\textbf{probe}}}$ is sent through the system and the corresponding intensity vector $\mathbf{I_u}$ is captured. By inverting Eq. (\ref{eq:3}), we obtain the reconstructed spectrum vector $\mathbf{S_u^{\textbf{reconstructed}}}$ based on the calibration data and the measured intensity,
\begin{equation}
\mathbf{ S_u^{\textbf{reconstructed}}} =  \mathbf{T^{-1}} * \mathbf{I_u}.
\label{eq:4}
\end{equation}
The performance of the spectrometer is measured by comparing the reconstructed spectrum to the actual probe spectrum, where the comparison is done by calculating the root mean square error,
\begin{equation}
\mu = \sqrt{ \frac{1}{M} [\mathbf{S_{\textbf{probe}}}  - \mathbf{ S_{\textbf{reconstructed}} }  ]^2 } .
\label{eq:5}
\end{equation}

\section*{Results}

\begin{figure*}[!h]
\centering \includegraphics[width=0.8\textwidth]{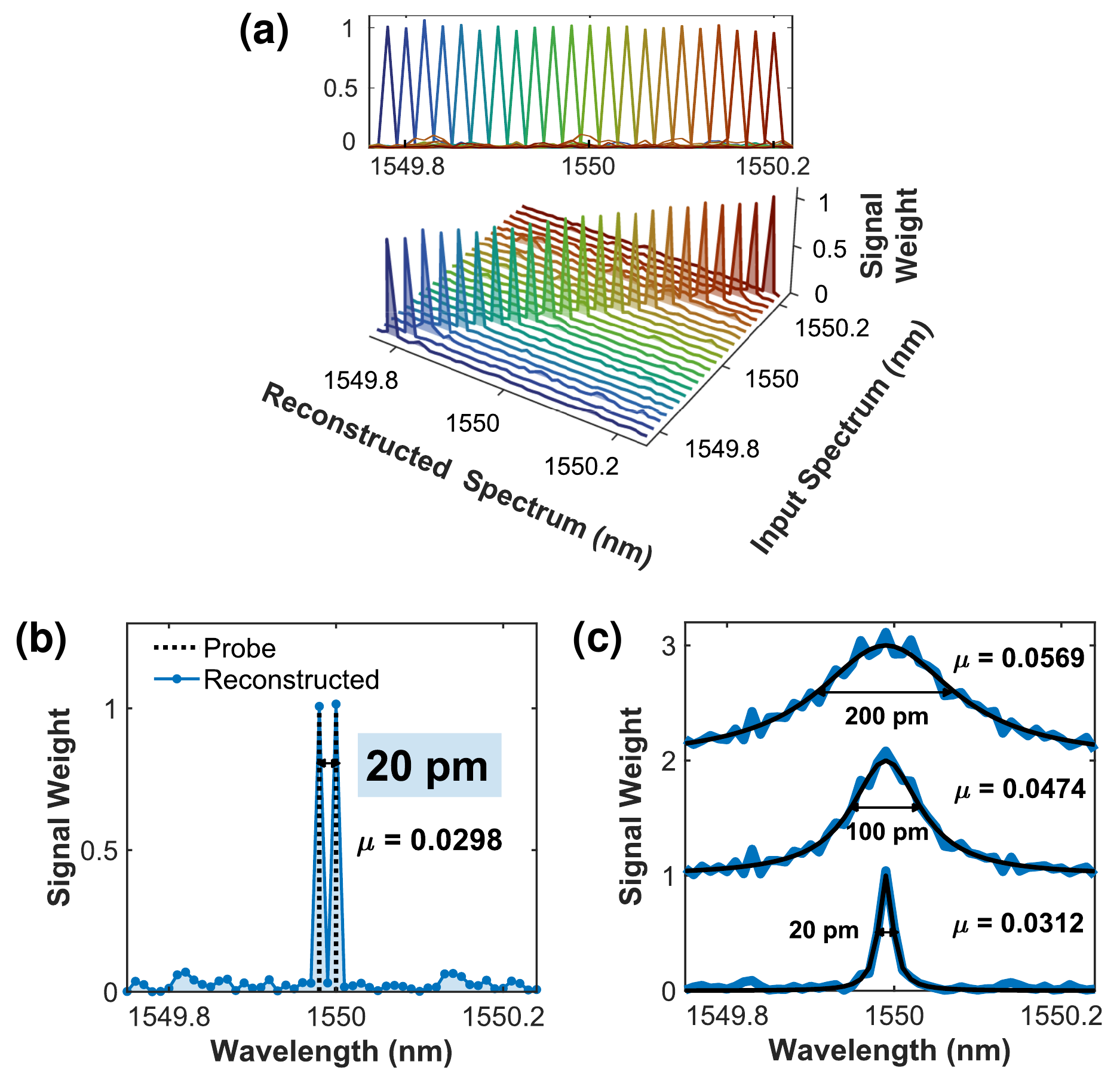}
\caption{Performance of spectrometer in reconstructing \textbf{(a)} individual wavelengths, \textbf{(b)} arbitrary spectra with two wavelengths, and \textbf{(c)} broadband Lorentzian spectra of various bandwidths.}
\label{fig:results}
\end{figure*}

Using the calibration process explained in Fig. \ref{fig:calibration}, we have calibrated our system with 50 wavelengths around the central wavelength 1550 nm (from 1549.75 nm to 1550.24 nm) with 10 pm step-size between consecutive wavelengths. After the calibration matrix is obtained and stored, individual wavelengths are sent through the system and measured intensities are plugged in Eq. (\ref{eq:4}). The reconstructed spectrum vectors of individual wavelengths in the spectral range are plotted in Fig. \ref{fig:results}(a). Since optical signals of different wavelengths do not interfere, an arbitrary spectrum can be modeled with a tunable laser by creating the spectrum content separately and then superposing the measured intensities. In the mathematical formulation, we sum the individual intensity vectors and plug the resulting vector in Eq. (\ref{eq:4}) as the measured intensity. For two wavelengths 20 pm apart from each other, the spectrum reconstruction is shown in Fig. \ref{fig:results}(b). We show that a single pixel spectrometer can resolve the spectral lines accurately and is able to reconstruct the unknown input spectrum with a reconstruction error $ \mu = 0.0298 $. 

While the input light entering the system may consist of sparse wavelengths, it may also carry a broadband spectrum. In this case, we model the continuous broadband spectrum as a weighted combination of individual wavelengths. We discretize the broadband spectrum with 10 pm spacing, which is the step size between two consecutive wavelengths in our dataset. To test the performance of the spectrometer, we modeled broadband Lorentzian beams of different bandwidths in the 490 pm spectral range from 1549.75 nm to 1550.24 nm. Each wavelength involved in the broadband signal is expected to contribute to the final intensity by its weight in the spectrum. Thus, the final intensity measured on the detector is found by superposing the individual intensity vectors scaled by the weight of the corresponding wavelength. Then, the resulting intensity vector is inserted in Eq. (\ref{eq:4}) to predict the input spectrum. The results for three Lorentzian beams centered at 1549.99 nm with 20 pm, 100 pm, and 200 pm bandwidths corresponding to $\sim$4\%, $\sim$20\% and $\sim$40\% of the total spectral range are plotted in Figure \ref{fig:results}(c). It is observed that the reconstruction error increases with the bandwidth of the signal, reaching $\sim$6\% ($\mu = 0.0569$) when $\sim$40\% of the spectral range is covered.

\section*{Discussion}

It is possible to reduce the spectrum reconstruction error of a multimode fiber based speckle spectrometer by suppressing the system noise using computational methods such as truncated inversion \cite{Redding2013}. In this study, we show pure physical results solely based on intensity measurements without any computational aid. The computational manipulations should be handled carefully in a single-pixel spectrometer since 1) the number of data points are fewer compared to a speckle spectrometer which may immediately dismiss some methods, and 2) diagonal data is positively biased in the calibration matrix of the single-pixel spectrometer opposite to the speckle spectrometer where the data is randomly distributed due to the nature of a speckle images. A computational method preserving the diagonality of the calibration matrix and simultaneously respecting the correlation between the columns of $\mathbf{T}$ may be a good candidate for reducing the errors. Increasing the isolation of the system by stabilizing the multimode fiber is another way of suppressing the system noise \cite{Wan2015}. In preferred cases, the isolation can be provided by replacing the multimode fiber with an integrated ridge waveguide which is much less susceptible to environmental variations. With such improvements, we believe the single-pixel spectrometer will be able to go beyond the resolution limit of a speckle spectrometer which is more affected by the detection noise \cite{Cao2017}.

\section*{Conclusion}

In summary, we have demonstrated a high-resolution, single-pixel multimode fiber spectrometer employing wavefront shaping of light. The working principle of the spectrometer is based on abrupt distortion of the focused intensity when the wavelength of the incoming light changes. Thus, the use of the SLM is not required for our method. The system can be externally perturbed by heat or carrier injection to a semiconductor waveguide. The calibration can be performed with respect to the external perturbation and an arbitrary spectrum can be reconstructed accordingly. The SLM, on the other hand, brings an additional advantage by increasing the signal-to-noise level. The efforts regarding the miniaturized  wavefront controllers can pave the way for simultaneous spectral reconstruction as well as increasing the SNR \cite{WeiHuang2020, Guo2021, Deng2022}. The proposed spectrometer is promising in developing single-pixel hyperspectral imaging applications across scattering media and it offers replacing bulky, expensive cameras with a single-pixel detector to develop low-budget systems.

{\large \textbf{Funding.}} TÜBİTAK (2211-A, 118M199). Turkish Academy of Sciences (TÜBA-GEBİP). 

{\large \textbf{Acknowledgements.}} We thank Alpan Bek for useful discussions. 

{\large \textbf{Data availability.}} Data underlying the results presented in this paper are publicly available at https://doi.org/10.5281/zenodo.7195860.


\end{document}

%% file: packages.tex
\usepackage[utf8]{inputenc}
\usepackage[english]{babel}
\usepackage[margin=1in, paperwidth=8.5in, paperheight=11in]{geometry}
\usepackage{amsmath, amssymb, url, relsize, soul, dashrule}
\usepackage[autostyle]{csquotes}
\MakeOuterQuote{"}
\usepackage{amsthm,bm}

\usepackage[font=small,labelfont=bf,margin=1cm]{caption}

\usepackage{pdfpages}

\usepackage{mathtools}

\usepackage{array,booktabs}
\newcolumntype{C}{>{$\displaystyle}c<{$}}

\usepackage{parskip} 
\usepackage{wrapfig} 
\usepackage{color} 

\usepackage{hyperref}

\usepackage{etoolbox}

\allowdisplaybreaks 

\usepackage[most]{tcolorbox}